\documentclass[useAMS,usenatbib]{mn2e}
\input epsf
\def\tem#1{\par\noindent
\hangindent6.5 mm\hangafter=0
\llap{#1\enspace}\ignorespaces}

\title[Evolutionary models of Star 32]{Evolutionary models of
the optical component of the LMC X-1/Star 32 binary system}
\author[J. Zi\'o{\l}kowski]{J. Zi\'o{\l}kowski \thanks{E-mail:
jz@camk.edu.pl}\\N. Copernicus Astronomical Center, ul. Bartycka
18, 00-716 Warsaw, Poland}
\begin{document}

\date{Accepted 0000 December 00. Received 0000 December 00; in original form 0000 December 00}

\pagerange{\pageref{firstpage}--\pageref{lastpage}} \pubyear{2009}

\maketitle

\label{firstpage}

\begin{abstract}
Calculations carried out to model the evolution of Star 32
 under different assumptions about the stellar wind mass-loss rate
provide robust limits on the present mass of the star. The obtained
range is 31 to 35.5 M$_\odot$, which is in very good agreement with
the orbital solution of Orosz et al., namely 28.3 to 35.3 M$_\odot$.
The initial mass of Star 32 had to be in the range 35 to 40
M$_\odot$ and the present age of the system is 3.7 to 4.0 Myr.

\end{abstract}


\section{Introduction}

LMC X-1 was one of the first X-ray binaries discovered, and the
first one found in Magellanic Clouds (Mark et al. 1969). The X-ray
source is persistent, although variable, the luminosity varying by
one order of magnitude (Liu, van Paradijs \& van den Heuvel 2005).
The source is very bright ($L_{\rm X} \sim 1 \div 2\times 10^{38}$
erg s$^{-1}$; Johnston, Bradt \& Doxsey 1979). Its X-ray spectrum
was found to be relatively soft ($kT \sim 2.7$ keV, Markert \& Clark
1975), which prompted Hutchings, Crampton \& Cowley (1983, hereafter
H83) to notice that it was similar to the two then known black hole
candidates: Cyg X-1 and LMC X-3. The optical identification was
firmly determined only quite recently (Cowley et al. 1995). The
difficulty with the identity of the optical counterpart was that
there were two comparably good candidates. Initially the bright ($V$
= 12.0) B5 supergiant R148 was favored as the counterpart (Jones,
Chetin \& Liller 1974; Rapley \& Tuohy 1974; Johnston et al. 1979).
The other, less favored candidate was the fainter ($V$ $\approx$
14.5) star denoted by Cowley et al. (1978) as star no. 32. This star
was observed spectroscopically by Pakull (1980), who noticed that
the spectra "show He II 468.6-nm and N III-C III 464-465-nm
emissions with strength comparable to that seen in most massive
X-ray binary systems". Both optical candidates were observed
spectroscopically by Hutchings et al. (1983). They found no
detectable periodic variability for R148, but radial velocity
measurements of Star 32 have shown it to be a binary with an orbital
period of approximately 4 days. Two equally good fits to the
observations were obtained for values of the period equal to 3.908
and 4.038 d. The authors also measured the velocities of emission
lines, discovered by Pakull, which were found to vary in antiphase
with the lines of Star 32. These lines probably originated near the
secondary (compact) component, and, from the relative amplitudes,
the authors deduced that the mass ratio (defined as the mass of the
optical to the mass of the compact component) should exceed 2. The
estimated most likely masses of the components were about $\sim$ 14
and $\sim$ 4 M$_\odot$, which suggested that the secondary may be a
black hole. White and Marshall (1984) noticed that the very soft
X-ray spectrum of LMC X-1 places it close to other black hole
candidates such as LMC X-3 and Cyg X-1 in an X-ray colour-colour
diagram. After further optical spectroscopy, Hutchings et al. (1987,
hereafter H87) refined the orbital period to 4.2288 d, and the
masses of the components to 20 and 6 M$_\odot$. The evidence that
Star 32 is indeed the optical counterpart of LMC X-1 looked quite
strong, but it was firmly established only after Cowley et al.
(1995) compared the position of Star 32 with the position of LMC X-1
from ROSAT observations. In 2006, Levine \& Corbet found the X-ray
orbital period from data from RXTE. Their period was equal to 3.9081
$\pm$ 0.0015 d, which was somewhat shorter than the then accepted
H87 period (4.2288 d), but was equal to one of the original (1983)
propositions of H83.

In 2005 I calculated the preliminary evolutionary tracks modelling
the evolution of Star 32 (Zi\'o{\l}kowski 2006). I found that the
present mass of this star should be in the range 24 to 33 M$_\odot$.
However, this result was difficult to reconcile with the then
available spectroscopic data. The reason was as follows.

We can estimate the radius $R_{\rm opt}$ of the optical component
from the observationally determined luminosity and the effective
temperature. For spectral type O7 III (H83) we have $T_{\rm e}
\approx 36000$ K and $B.C. \approx -$3.50. H83 estimated reddening
as $E_{\rm B-V} \approx 0.35$, which implies $A_{\rm V} \approx
1.1$. With $V \approx$ 14.8 and the distance modulus to the Large
Magellanic Cloud (LMC) equal to 18.5 we have $M_{\rm bol} \approx
-8.3$ or log($L$/L$_{\odot}$) $\approx$ 5.2. With this luminosity
and effective temperature, one obtains $R_{\rm opt} \approx 10.5
R_{\odot}$.

If we assume the mass of the optical component $M_{\rm opt}$ to be a
free parameter, then after selecting its value we can use two
equations to solve for the inclination of the orbit $i$ and the mass
ratio $q = M_{\rm opt}/M_{\rm x}$ (where $M_{\rm x}$ is the mass of
the compact component). One of these equations makes use of the mass
function:
\begin{eqnarray}
f(M_{\rm x}) = M_{\rm opt} {\rm sin}^3 i/[q(1+q)^2].
\end{eqnarray}

\noindent The other relates the radius of the star to the size of
the orbit:
\begin{eqnarray}
R_{\rm opt}& =& R_{\rm RL} \times f_{\rm RL} =\nonumber\\ &=& f_{\rm
RL} (0.38 + 0.2 \log\hspace*{.5ex} q) A =\nonumber \\ &=& f_{\rm RL}
(0.38 + 0.2 \log\hspace*{.5ex} q) a_1 (1+q),
\end{eqnarray}

\noindent where $R_{\rm RL}$ is the radius of the Roche lobe (e.g.
Paczy\'nski 1971) around Star 32, $f_{\rm RL}$ is the fill-out
factor ($f_{\rm RL} = R_{\rm opt}/R_{\rm RL}$), $A$ is the orbital
separation of the binary components and $a_1$ is the radius of the
orbit of Star 32.

Adopting, after H87, the (then accepted) values of the orbital
period (4.2288 d) and of the radial velocities amplitude ($K_{\rm
opt}$ = 69 km/s), one has $f(M_{\rm x})$ = 0.144M$_\odot$ and
$a_1$sin$\hspace*{.5ex} i$ = 5.76R$_\odot$ for the mass function and
the radius of the orbit of the optical component, respectively.

Inserting these data, equations (1)$-$(2) can be written as
\begin{eqnarray}
M_{\rm opt} {\rm sin}^3 i/[q(1+q)^2] = 0.144,
\end{eqnarray}
\begin{eqnarray}
R_{\rm opt} = f_{\rm RL} (0.38 + 0.2 \log \hspace*{.5ex} q) (1+q)
\times 5.76/{\rm sin}\hspace*{.5ex} i.
\end{eqnarray}

Recall that both H83 and H87 indicate that the surface of the star
must be near the Roche limiting surface ($f_{\rm RL} \ga 0.9$).
Then, solving equations (3)$-$(4) for $i$ and $q$, it is easy to
verify that the observational condition $q \ga 2$ (H83) can be
satisfied only for $M_{\rm opt} \la 8$ M$_\odot$ if $f_{\rm RL}$ =
0.95 or for $M_{\rm opt} \la 9.5$ M$_\odot$ if $f_{\rm RL}$ = 0.9.
Allowing for observational errors, one might increase these upper
limits slightly for the value of $M_{\rm opt}$, but this would not
bridge the gap between the orbital solution ($M_{\rm opt} \la 8 \div
9$ M$_\odot$, as shown above) and the evolutionary constraints
($M_{\rm opt} \ga 24$ M$_\odot$).

Therefore, I concluded my 2006 considerations with the statement
"There are only two possible solutions of this discrepancy: either
observations of Star 32 (spectroscopy and/or photometry) are in
serious error or Star 32 is essentially a helium star (with only a
small remnant of the hydrogen rich envelope)". I concluded that
future observations should solve this problem.

\section{New Orbital Solution}

Later observations indeed solved the problem. Orosz et al. (2009)
published a detailed analysis of their new extensive spectroscopic
and photometric observations of Star 32 together with a refined
analysis of the ASM data from RXTE observations of LMC X-1. Their
new orbital solution greatly improved the earlier rough estimates,
removed the puzzles (mass versus luminosity of the optical
component) and produced a fairly consistent picture of the binary
system. First, they found that the optical orbital period is not
4.2288 d (as was generally accepted after H87), but rather 3.90917
$\pm$ 0.00005 d, in agreement with the X-ray period suggested by
Levine \& Corbet (2006), and also with one of the two original
orbital period candidates of H83. After a thorough analysis, they
determined the masses of the components as $M_{\rm opt} = 30.62 \pm
3.22$ M$_\odot$ and $M_{\rm X} = 10.30 \pm 1.34$ M$_\odot$. They
also determined quite precisely the luminosity and the effective
temperature of Star 32 as log($L$/L$_{\odot}$) = 5.50 $\pm$ 0.05 and
$T_{\rm e}$ = 33200 $\pm$ 500 K. They found that these values are
consistent with the core burning star of initial mass $\sim$ 35
M$_\odot$.

It thus became clear that the reasons for the dramatic discrepancy
discussed in Zi\'o{\l}kowski (2006) were: (i) a serious (by a factor
of 2) underestimate of the luminosity of Star 32 and (ii) an
overestimate (by $\sim 8.5$ per cent) of its effective temperature.
These two factors led to a serious (by $\sim 40$ per cent)
underestimate of its radius ($\sim 10.5$ R$_\odot$, instead of $\sim
17$ R$_\odot$). This small value for the radius of Star 32, together
with the requirement that it should nearly fill its Roche lobe, led
to an uncomfortably small mass of this star.

Note that even the underestimated luminosity of log($L$/L$_{\odot}$)
$\approx$ 5.2 was much too high for the mass of Star 32 proposed by
H83 ($\sim$ 14 M$_\odot$; and this relatively high value of the mass
was achieved only by stretching the then observed value of radius
from $\sim 10$ R$_\odot$ to $\sim 12$ R$_\odot$).

After new precise data concerning the LMC X-1/Star 32 binary system
became available, I decided to carry out new evolutionary
calculations to determine more precisely the past evolution and the
present evolutionary state of Star 32.

\section{Evolutionary Calculations for Star 32}

\subsection{General description}

I computed evolutionary tracks for core hydrogen-burning phase of
stars with initial masses in the range $30$--$50$ M$_\odot$. The
Warsaw evolutionary code developed by Bohdan Paczy\'nski and Maciek
Koz{\l}owski and updated by Ryszard Sienkiewicz and Alosha
Pamyatnykh was used. An initial chemical composition of $X$=0.7 and
$Z$=0.008, appropriate for LMC, was adopted. Opacity tables
incorporating OPAL opacities (Iglesias \& Rogers 1996) as well as
molecular and grain opacities (Alexander \& Ferguson 1994) were
used. The nuclear reaction rates are those of Bahcall \&
Pinsonneault (1995). The equation of state used was that of
Livermore Laboratory OPAL (Rogers, Svenson \& Iglesias 1996).
Semiconvective mixing was neglected, as it is not important during
the evolutionary phase considered (most of the models of interest
had central hydrogen content $X_{\rm c} \ga 0.2$). Similarly, any
overshooting at the border of the convective core was neglected, as
this is even less important.

The calculations were carried out under the assumption that the
evolution starts from homogeneous configurations. This means that
the consequences of the fact that some of the matter of the star,
possibly dumped from the progenitor of the present black hole, could
have modified the chemical composition, were neglected. It also
means that the consequences of the fact that some nuclear evolution
(hydrogen burning) could, possibly, have taken place while the mass
of the star was smaller (prior to the mass transfer) were neglected
as well. It seems that neither simplification significantly alters
the outcome of the evolutionary calculations. The orbital period is
so short that any substantial mass transfer during past evolution
seems unlikely, as it would have lead to a common envelope
configuration and the merger of the two stars. Even if (which seems
unlikely) there was significant mass transfer in the system, then it
just reset the evolutionary clock and we can consider the evolution
of the mass gainer as starting anew from the zero-age main sequence
(ZAMS), as a single star but with a higher mass.

\subsection{Stellar wind mass loss}

The most uncertain element of the calculations of the early
evolution of massive stars is the mass loss caused by the stellar
wind. The uncertainty in the estimate of its rate is the single most
important factor influencing the outcome of the calculations (see
e.g. Zi\'o{\l}kowski 2005). The observations seem to indicate that
there is a substantial scatter of mass-loss rate among stars of
similar luminosities and effective temperatures. The commonly used
formula derived by Hurley, Pols \& Tout (2000, hereafter HPT), based
on parametrization of Nieuwenhuijzen \& de Jager (1990), gives the
estimate of the mass-loss rate with an accuracy that is probably not
better than a factor of two. Bearing this in mind, I introduced the
multiplicative factor $f_{\rm SW}$ applied to the HPT formula. In
this way, the uncertainty in the theory of evolution could be, in
some way, taken into account.

The value of the factor $f_{\rm SW}$, which should be used to model
successfully the evolution of a given star, might be quite different
for seemingly similar stars. For some cases (e.g. HDE 226868, which
is the companion to Cyg X-1), this value is close to 1
(Zi\'o{\l}kowski 2005). However, in some high mass X-ray binaries,
the optical supergiant components are significantly under-massive
for their luminosities (Zi\'o{\l}kowski 1977). In some systems, such
as Cen X-3, this undermassivness is very serious and requires much
stronger mass loss than the normal HPT stellar wind (Zi\'o{\l}kowski
1978). It seems that Star 32 is probably (as we shall see) closer to
the case Cen X-3 than to HDE 226868, in this respect.

The evolutionary calculations carried out to produce models
reproducing the present evolutionary state of Star 32 and using the
HPT formula are quite successful (see Fig. 1a), except for one
aspect: the value of the stellar wind mass-loss rates predicted by
these models are too small by almost an order of magnitude when
compared with the observations (see Table 1). The observed value
$\dot{M} \approx -5 \times 10^{-6}$ M$_\odot$/yr was determined by
Orosz et al. (2009) from analysis of the orbital X-ray light curve.
They modelled the light curve successfully, assuming that
variability is due to the scattering of X-ray photons by electrons
in the stellar wind over the variable (with the orbital phase)
optical depth in the wind. This estimate agrees roughly with the one
derived from the observed X-ray luminosity. It should be noted,
however, that this estimate, being model dependent, is not as robust
as the determinations of the parameters of the system and its
uncertainty might be quite large.

The observed value of $\dot{M}$ seems to be substantially (by a
factor of $\sim$ 6 -- 8) larger than the value predicted by the HPT
formula for the present parameters of Star 32. It is clear,
therefore, that to model the evolution of Star 32 (including its
present mass-loss rate) successfully it is necessary to modify the
HPT formula, increasing significantly the strength of the stellar
wind; that is, to use values of the factor $f_{\rm SW}$ that are
substantially larger than 1. I have tried three such modifications:

\begin{eqnarray}
f_{\rm SW} = f_{\rm now},
\end{eqnarray}
\begin{eqnarray}
f_{\rm SW} = 1+(f_{\rm now}-1)(R-R_{\rm ZAMS})/(R_{\rm cr}-R_{\rm
ZAMS}),
\end{eqnarray}
\begin{eqnarray}
f_{\rm SW} = 1+(f_{\rm now}-1)[(R-R_{\rm ZAMS})/(R_{\rm cr}-R_{\rm
ZAMS})]^2,
\end{eqnarray}

In these equations, $f_{\rm now}$ denotes the value of $f_{\rm SW}$
corresponding to the present state of Star 32, that is, the value
one has to apply to reconcile the calculated mass flux of the
stellar wind (for the evolutionary model corresponding to the
present state of Star 32) with the observed one. It was found to be
$\sim$ 6 for the upper edge and $\sim$ 8.5 for the lower edge of the
observational error box (the crosses in Figs 1a--d). $R$ is the
temporary stellar radius, $R_{\rm ZAMS}$ is the initial value of the
stellar radius (on the ZAMS) and $R_{\rm cr}$ is the present radius
of the inner Roche lobe around the optical component.

\begin{table}

\centering
 \begin{minipage}{75mm}
  \caption{Parameters of selected evolutionary models for Star 32}

\vbox{
\begin{tabular}{|c|l|c|rcl|c|c|}
\hline

\multicolumn{1}{|c|}{No.}&\multicolumn{1}{|c|}{$M_0$}&\multicolumn{1}{|c|}{$M_{\rm opt}$}&\multicolumn{3}{|c|}{$f_{\rm now}$}&\multicolumn{1}{|c|}{log$L$}&\multicolumn{1}{|c|}{$-\dot{M}$}\\
&\multicolumn{1}{|c|}{[M$_{\odot}$]}&\multicolumn{1}{|c|}{[M$_{\odot}$]}&&&&\multicolumn{1}{|c|}{[L$_{\odot}$]}&
\multicolumn{1}{|c|}{[10$^{-6}$ M$_{\odot}$/yr]}\\

\hline
&&&&&&&\\

1&33.5&32.34&1\hspace*{-3.5ex}&&&5.446&0.57\\
2&37.5&35.93&1\hspace*{-3.5ex}&&&5.545&0.84\\
3&41.5&29.66&8\hspace*{-3.5ex}&.&\hspace*{-3.5ex}5&5.451&4.85\\
4&44.5&33.66&6\hspace*{-3.5ex}&&&5.551&5.10\\
5&36&31.28&8\hspace*{-3.5ex}&.&\hspace*{-3.5ex}5&5.443&4.66\\
6&40&35.19&6\hspace*{-3.5ex}&&&5.550&5.07\\
7&35&31.79&8\hspace*{-3.5ex}&.&\hspace*{-3.5ex}5&5.446&4.63\\
8&39&35.58&6\hspace*{-3.5ex}&&&5.550&5.04\\

\end{tabular}}

\vspace{4mm}
{\footnotesize NOTES:\vspace{2mm}\\
(i) No. denotes the number of the evolutionary sequence; sequences 1
and 2 were obtained using the HPT formula (see text); sequences 3
and 4 used formula (5); sequences 5 and 6 used formula (6); and
sequences 7 and 8
used formula (7)\vspace{1mm}\\
(ii) $M_0$ denotes the initial (zero-age main sequence) mass of the
optical component; $M_{\rm opt}$, the mass at the evolutionary phase
corresponding to the present state of Star 32; $f_{\rm now}$, the
multiplicative factor applied in formulae (5)--(7); other symbols
have their usual meanings\\} \vspace{5mm}
\end{minipage}
\end{table}

The modification given by equation (5) is the simplest one, but is
not very realistic. There is no reason why the stellar wind should
be, from the very beginning, stronger than the typical one (HPT) by
a factor as high as 6--8. However, there are indications, dating
from a long time ago (Zi\'o{\l}kowski 1977, Hutchings 1979), that
the strength of the stellar wind from a component of a close binary
system might increase significantly as the stellar surface
approaches

\begin{figure*}
\begingroup
   \def \A#1{\epsfxsize=0.48\hsize \epsfbox{#1}}
   \hbox to \hsize{\A{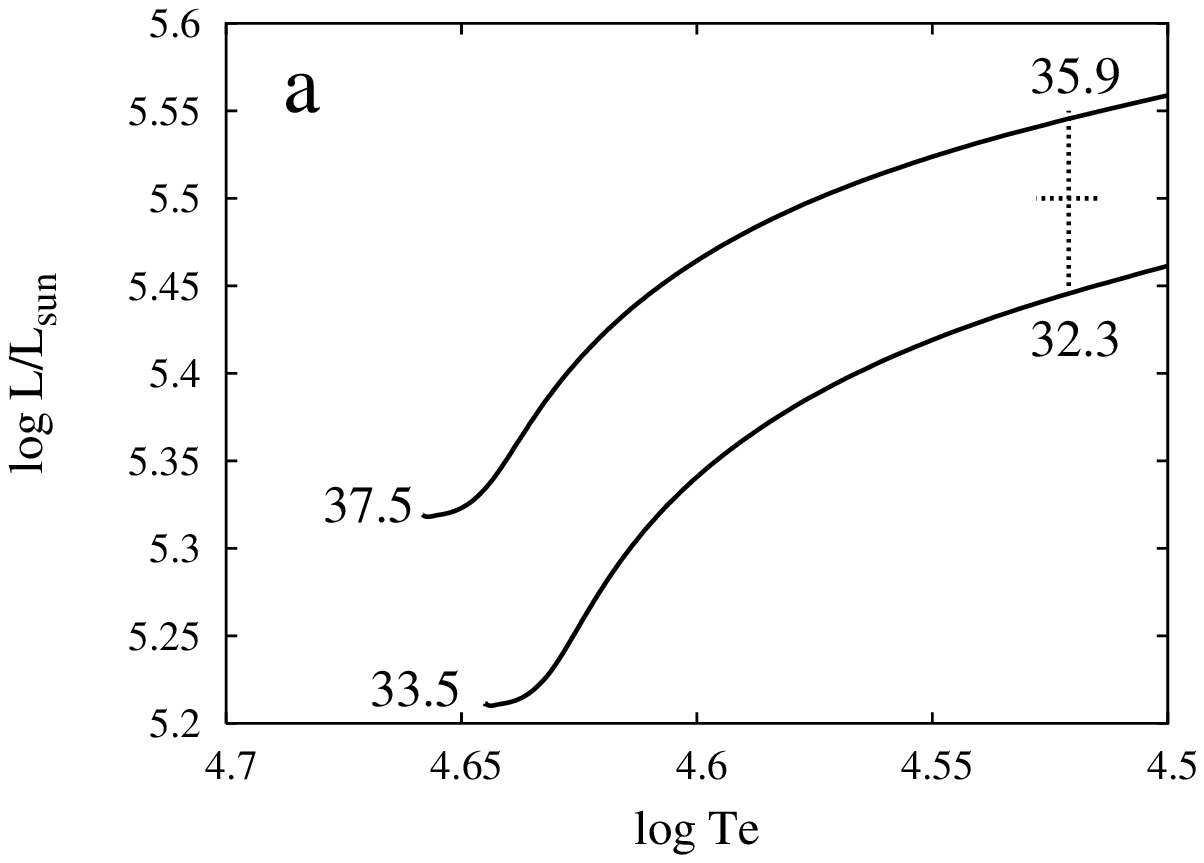}\hfil\A{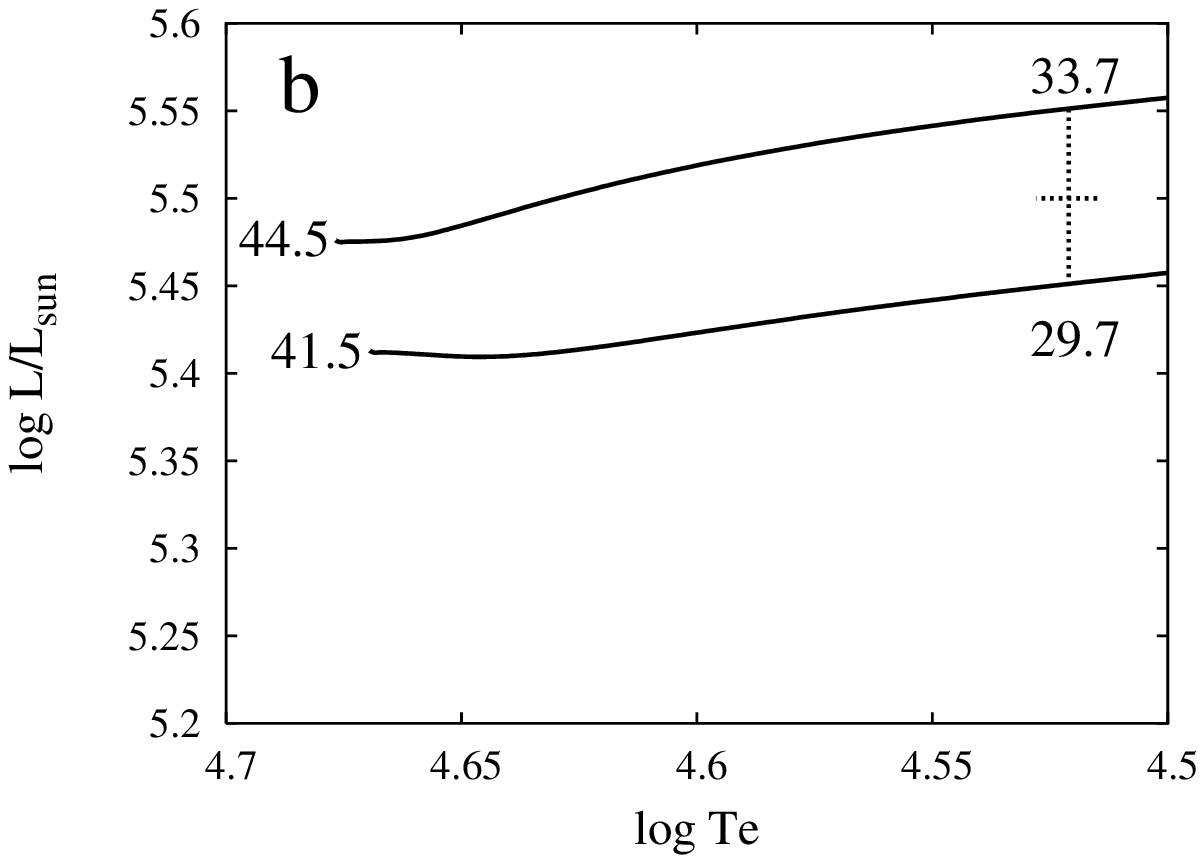}}
   \hbox to \hsize{\A{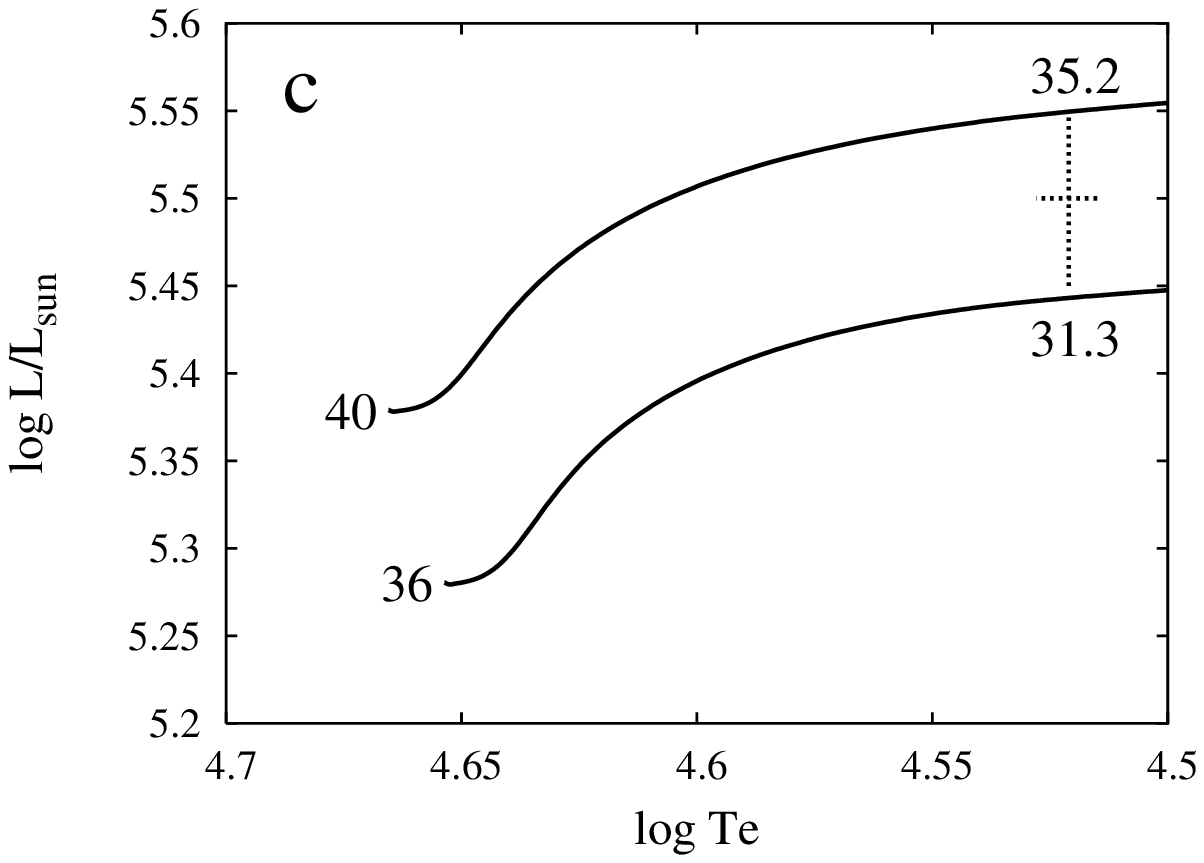}\hfil\A{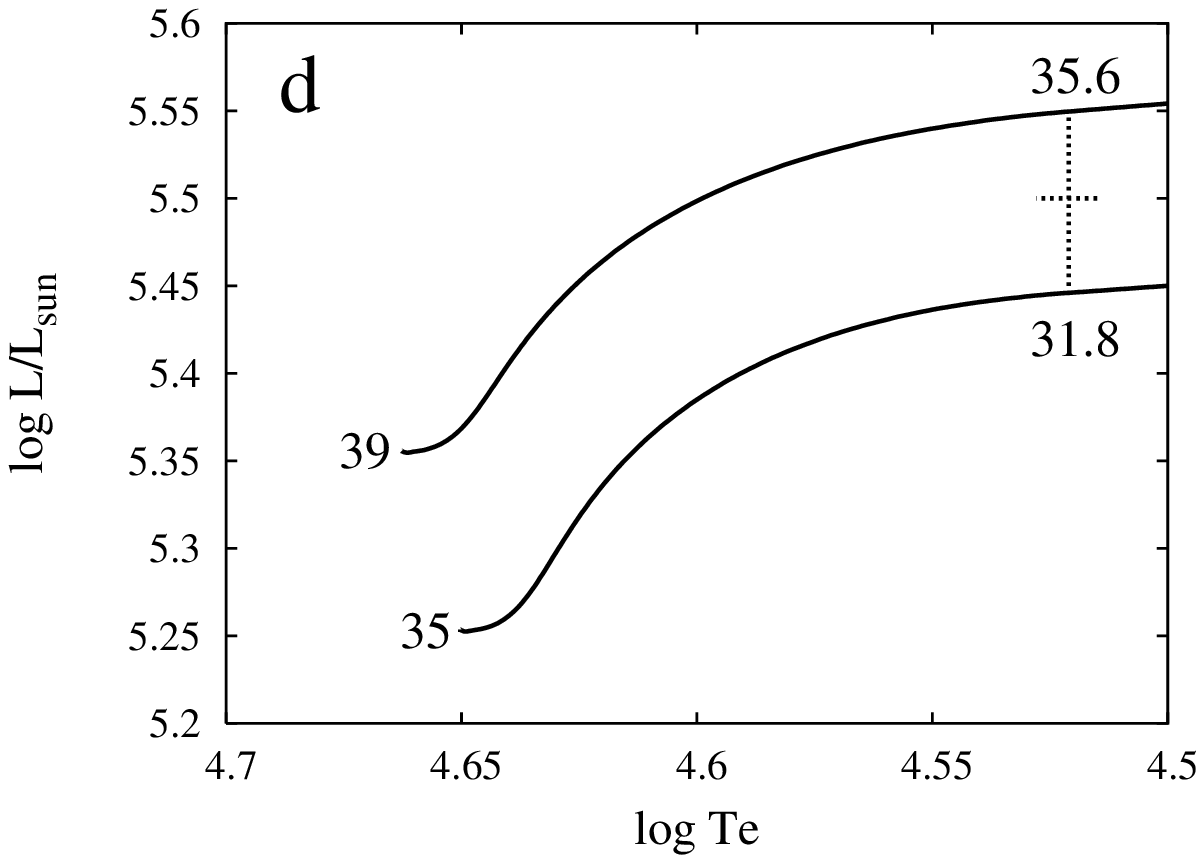}}
\endgroup
      \caption[h]{The evolutionary tracks in the Hertzsprung-Russell diagram (solid lines).
The dotted cross shows the observed position of Star 32. The number
at the start of each track indicates the initial mass (in solar
units). The numbers near the cross indicate the mass at the
evolutionary phase corresponding to the present state of Star 32.
Part (a) shows the tracks obtained using the HPT formula to describe
the stellar wind (see text); (b) shows tracks obtained using formula
(5); (c) shows tracks obtained using formula (6); and (d) shows
tracks obtained using formula (7). The value of the parameter
$f_{\rm now}$ (see text) used for each track is given in Table 1.}
     \label{f1}
    \end{figure*}

\noindent that of the Roche lobe. There is no quantitative
description of this effect available. The formulae (6) and (7) try
to incorporate it, in a crude way, by a linear or quadratic
dependence on the distance from the Roche lobe.

\subsection{Results}

The evolutionary tracks in the Hertzsprung-Russell (HR) diagram,
obtained with the unmodified HPT formula and with formulae (5)--(7),
are shown in Figs 1(a)--(d). All tracks reproduce the present
luminosity and effective temperature of Star 32 well, and the tracks
obtained with formulae (5)--(7) reproduce, in addition, the present
mass flux of the stellar wind. The parameters of the models
corresponding to the present state of Star 32 are given in Table 1.
The variations of the stellar wind mass flux with the evolutionary
time, for different evolutionary sequences, are shown in Fig. 2.

The summary of the results can be brief. Equation (5) does not seem
to correspond to the true history of the stellar wind strength in
LMC X-1/Star 32 binary system (see the discussion in Section 3.2).
The formulae (6) and (7) (although obtained ad hoc) seem to be much
more realistic, and I believe that they give a crude but reasonable
description of the evolution of the stellar wind mass flux from Star
32. The differences between the results obtained with formula (6)
and formula (7) are not significant (see Table 1 and Figs 1c and
1d), which makes our conclusions more robust. To conclude: the
present {\it evolutionary} mass of Star 32 must be in the range 31
to 35.5 M$_\odot$. This result is in very good agreement with the
new orbital solution of Orosz et al. (2009), who give a range for
the mass of the star  of 28.3 to 35.3 M$_\odot$. According to our
evolutionary sequences, the initial mass of Star 32 had to be in the
range 35 to 40 M$_\odot$ and the present age of the system is 3.7 to
4.0 million Myr.

\begin{figure}
\hbox{\epsfxsize=1\hsize\epsfbox{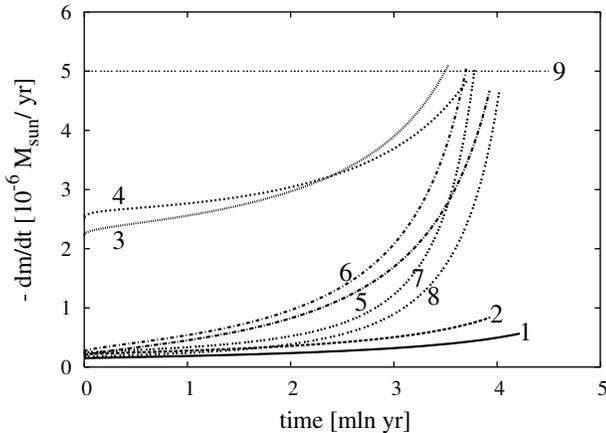}}
      \caption[h]{The stellar wind mass loss-rate as a function of
      evolutionary time. Each sequence is labelled with the number
      corresponding to that in Table 1. The line no. 9 describes
      (approximately) the rate observed now for Star 32.}
     \label{f2}
    \end{figure}

In addition, note that if one disregards the estimate of $\dot{M}$,
as a not very reliable constraint, then one is left with sequences 1
and 2. The present mass of Star 32 is then in the range 32 to 36
M$_\odot$, its initial mass is in the range 33.5 to 37.5 M$_\odot$,
and the present age of the system is in the range 3.9 to 4.2 Myr. As
can be seen, the results are not very different. This confirms that
our conclusions are not sensitive to the uncertainties of the
mass-loss treatment.

\section{Conclusions}

\tem{(i)} The calculations modelling the evolution of Star 32, under
different assumptions about the shape of the formula describing the
stellar wind mass-loss rate, provide robust limits on the present
mass of the star. For the reasonable versions of this formula, this
mass should be in the range 31 to 35.5 M$_\odot$.

\tem{(ii)} The the initial mass of Star 32 had to be in the range 35
to 40 M$_\odot$.

\tem{(iii)} The present age of the system (counted from its birth or
from the resetting of the evolutionary clock after the possible, but
unlikely, mass transfer) is in the range 3.7 to 4.0 Myr.

\tem{(iv)} The evolutionary estimate of the mass of Star 32 remains
in very good agreement with the estimate based on the orbital
solution of Orosz et al. (2009): 28.3 to 35.3 M$_\odot$.

\section*{Acknowledgements}

I would like to thank A. Zdziarski for a careful reading of the
manuscript and for many helpful comments and stimulating
discussions. I would like also to thank the referee, Ian Howarth
whose comments helped to improve this paper significantly. This work
was partially supported by the Polish Ministry of Science and Higher
Education project 362/1/N-INTEGRAL (2009-2012).

\label{lastpage}
\end{document}